\documentstyle[12pt]{article}

\def\d{\delta}
\def\e{\epsilon}

\def\f{\phi}
\def\vf{\varphi}
\def\g{\gamma}

\def\j{\psi}
\def\k{\kappa}
\def\l{\lambda}
\def\m{\mu}

\def\q{\theta}

\def\x{\xi}
\def\z{\zeta}
\def\D{\Delta}
\def\F{\Phi}
\def\G{\Gamma}

\def\Ome{\Omega}
\def\Q{\Theta}
\def\S{\Sigma}

\def\X{\Xi}

\def\cf{{\cal F}}

\def\co{{\cal O}}

\def\cs{{\cal S}}

\def\inbar{\vrule height1.5ex width.4pt depth0pt}
\def\rlx{\relax\leavevmode}
\def\I{\leavevmode\hbox{\small1\kern-3.8pt\normalsize1}}
\def\openone{\leavevmode\hbox{\small1\kern-3.3pt\normalsize1}}
\def\Ione{\rlx{\rm 1\kern-2.7pt l}}
\def\Ik{\rlx{\rm I\kern-.18em k}}  
\def\IC{\rlx\leavevmode
	     \ifmmode\mathchoice
		    {\hbox{\kern.33em\inbar\kern-.3em{\rm C}}}
		    {\hbox{\kern.33em\inbar\kern-.3em{\rm C}}}
		    {\hbox{\kern.28em\sinbar\kern-.25em{\rm C}}}
		    {\hbox{\kern.25em\ssinbar\kern-.22em{\rm C}}}
	     \else{\hbox{\kern.3em\inbar\kern-.3em{\rm C}}}\fi}
\def\IP{\rlx{\rm I\kern-.18em P}}
\def\IR{\rlx{\rm I\kern-.18em R}}
\def\IN{\rlx{\rm I\kern-.20em N}}

\def\llsymbol#1{\@llsymbol{\@nameuse{c@#1}}}
\def\@llsymbol#1{\ifcase#1\or {}\or {'}\or {''}\or {'''}\or
   {''''}\or {'''''}\or  \else\@ctrerr\fi\relax}

\newcounter{contador}
\newcommand{\letra}{
   \stepcounter{equation}
   \setcounter{contador}{\value{equation}}
   \setcounter{equation}{0}
   \renewcommand{\theequation}{\thecontador.\alph{equation}}}
\newcommand{\antiletra}{
   \renewcommand{\theequation}{\arabic{equation}}
   \setcounter{equation}{\value{contador}}}

\newcommand{\ol}\overline
\newcommand{\ti}\tilde
\newcommand{\wt}\widetilde
\newcommand{\wh}\widehat
\newcommand{\bv}\breve
\newcommand{\dg}\dagger
\newcommand{\pari}{\stackrel{{P}}\longrightarrow}

\newcommand{\aand}{\;\;\;\mbox{and}\;\;\;}

\newcommand{\be}{\begin{equation}}
\newcommand{\ee}{\end{equation}}
\newcommand{\bl}{\begin{eqnarray}&}
\newcommand{\el}{&\end{eqnarray}}
\newcommand{\bq}{\begin{eqnarray}}
\newcommand{\eq}{\end{eqnarray}}

\newcommand{\ov}{\overline}
\newcommand{\pa}{\partial}

\def\sl#1{\rlap{\hbox{$\mskip 1 mu /$}}#1}      
\def\Sl#1{\rlap{\hbox{$\mskip 3 mu /$}}#1}      
\def\SL#1{\rlap{\hbox{$\mskip 4.5 mu /$}}#1}    

\def\ssl#1{\rlap{\hbox{$ {\scriptstyle /}$}}#1}

\begin{document}

\title{\Large \bf Algebraic Renormalization of Parity-Preserving 
QED$_{3}$ Coupled to Scalar Matter II: Broken Case }

\author{{\it O. M. Del Cima} {\thanks{E-mail: 
oswaldo@cbpfsu1.cat.cbpf.br .}}~,~
{\it D. H. T. Franco} {\thanks{E-mail:
franco@cbpfsu1.cat.cbpf.br .}}~,~ {\it J. A. 
Helay\"el-Neto}\\
and~{\it O. Piguet} {\thanks{On leave of absence from 
{\it D\'epartement de Physique Th\'eorique - Universit\'e de Gen\`eve, 
24 quai E. Ansermet - CH1211 - Gen\`eve 4 - Switzerland}. E-mail:
piguet@sc2a.unige.ch .}} {\thanks{Supported in part by the {\it Swiss 
National Science Foundation}.}}\\$\,$\\
{\normalsize Centro Brasileiro de Pesquisas F\'\i 
sicas (CBPF)} \\
{\normalsize Departamento de Teoria de Campos e 
Part\'\i culas (DCP)}\\
{\normalsize Rua Dr. Xavier Sigaud, 150 - Urca} 
\\
{\normalsize 22290-180 - Rio de Janeiro - RJ -
 Brazil.}\\$\,$\\
{\normalsize UGVA--DPT--1996--10--953}}

\date{}

\maketitle

\begin{abstract}
		   
In this letter the algebraic renormalization method, which is independent 
of any kind of regularization scheme, is presented for the parity-preserving 
QED$_{3}$ coupled to scalar matter in the broken regime, where the scalar 
assumes a finite vacuum expectation value, ${\langle}\vf{\rangle}$$=$$v$. 
The model shows to be stable under radiative corrections and anomaly free.

\end{abstract}

In the present letter, the model proposed in ref.~{\cite{e-pair}} is 
renormalized, in the broken regime (where the scalar assumes a nonvanishing 
vacuum expectation value), by using the algebraic renormalization 
method~{\cite{brs,troisieme,sopi}}. This 
algebraic approach is based on the BRS-formalism~{\cite{brs}} 
together with the Quantum Action Principle~{\cite{qap}}, which leads to 
a regularization independent scheme. The stability of the model under 
radiative corrections is analyzed as well as the possible presence 
of anomalies. 
The algebraic renormalization of the model in the symmetric regime 
was presented in ref.~{\cite{arqed31}}.  

The gauge invariant action for the parity-preserving QED$_{3}$ 
coupled to scalar matter~{\cite{e-pair,arqed31}} is given by :
\bq
\S_{\rm inv}\!\!\!\!&=&\!\!\!\!\int{d^3 x}      
\left\{ -{1\over4}
F^{mn}F_{mn}
+ i {\ov\j _+} {\SL{D}} {\j}_+ + i
{\ov\j _-} {\SL{D}} {\j}_- - m_0(\ov\j_+\j_+ - 
\ov\j_-\j_-) \;+\right.   
\nonumber\\
&&\left.
 -\;y (\ov\j_+\j_+ - 
\ov\j_-\j_-)\vf^*\vf + D^m\vf^* D_m\vf - \m^2\vf^*\vf - 
{\z\over2}(\vf^*\vf)^2 -{\l\over3}(\vf^*\vf)^3 \right\} 
\;\;\;\;\;\;\;,
\label{qed3}
\eq
where the mass dimensions of the parameters $m_0$, $\m$, $\z$, 
$\l$ and $y$ are
respectively ${1}$ ,${1}$, ${1}$, ${0}$ and ${0}$.  

The covariant derivatives are defined as follows :
\be
{\SL{D}}\j_{\pm}\equiv(\sl{\pa} + iqg \Sl{A})\j_{\pm} 
\aand
D_{m}\vf\equiv(\pa_{m} + iQ g A_{m})\vf \;\;\;, 
\label{covder}
\ee
where $g$ is a coupling constant with dimension of 
(mass)$^{1\over2}$ and $q$
and $Q$ are the $U(1)$-charges of the fermions and 
scalar, respectively. In the
action (\ref{qed3}), $F_{mn}$ is the usual field
strength for $A_m$, $\j_+$ and $\j_-$ are two kinds 
of fermions (the $\pm$
subscripts refer to their spin sign~{\cite{binegar}}) 
and $\vf$ is a complex
scalar.

The form of the potential is chosen such as to ensure 
the broken regime, where ${\langle}\vf{\rangle}$$=$$v$. 
Imposing that it must be bounded from below and yields only stable 
vacua, we get the following conditions on the parameters :
\be
\l>0 \;\;,\;\;\; \z<0 \aand \m^2 \leq {3\over 16} 
{\z^2\over \l} \;\;\;.
\label{cond}
\ee 
The vacuum expectation value for
the $\vf^*\vf$-product, $v^2$, is chosen as the solution
\be
{\langle}\vf^*\vf{\rangle}=v^2=-{\z \over 2\l}+ 
\left[ \biggl({\z \over
2\l}\biggr)^2 - {\m^2\over \l} \right]^{1\over 2} 
\;\;\;, \label{vac}
\ee
of the equation
\be
\m^2+{\z}v^2+{\l}v^4=0 
\label{mincond}
\ee
expressing the minimization of the potential.
The complex scalar $\vf$ is parametrized by
\be
\vf=v+H+i\q\;\;\;, \label{para}
\ee
where $\q$ is the would-be Goldstone boson and $H$ 
is the Higgs scalar, both
with vanishing vacuum expectation values. It should be noticed 
that the parametrization given by Eq.(\ref{para}) does not introduce
nonrenormalizable interactions, to the contrary of the unitary gauge 
parametrization~\cite{abers}.

With the parametrization (\ref{para}), the action (\ref{qed3}) reads:
\bq
\S_{\rm inv}\!\!\!\!&=&\!\!\!\!\int{d^3 x}      
\left\{ -{1\over4}
F^{mn}F_{mn}
+ i {\ov\j _+} {\SL{D}} {\j}_+ + i
{\ov\j _-} {\SL{D}} {\j}_- - m_0(\ov\j_+\j_+ - 
\ov\j_-\j_-) \;+\right.   
\nonumber\\
&&\left.
 -\;y (\ov\j_+\j_+ - 
\ov\j_-\j_-)((v+H)^2+\q^2) + \pa^m H \pa_m H +
{\pa^m}\q {\pa_m}\q \;+\right.   
\nonumber\\
&&\left. 
+\;2vQgA^m{\pa_m}\q + 2QgA^m(H{\pa_m}\q - \q{\pa_m}H)+ Q^2g^2 A^m 
A_m((v+H)^2+\q^2)\;+
\right.   \nonumber\\   
&&\left. 
- \;\m^2((v+H)^2+\q^2) - {\z\over2}((v+H)^2+\q^2)^2 -
 {\l\over3}((v+H)^2+\q^2)^3 \right\} 
\;\;\;\;\;\;\;.
\label{inv}
\eq  
 
The masses arising from the action (\ref{inv}) for $\j_{\pm}$, 
$A_m$ and $H$, are respectively
given by $m$$=$$m_0+yv^2$, $M^2_A$$=$$2v^2Q^2g^2$ and 
$M^2_H$$=$$2v^2(\z+2 \l v^2)$.  

In order to quantize the system (\ref{inv}) one has to add a
gauge-fixing action $\S_{\rm gf}$ -- we choose the $\x$-gauge --
and an action term $\S_{\rm ext}$ for the coupling of the BRS 
transformations to external sources :
\bq
\S_{\rm gf}\!\!\!\!&=&\!\!\!\!\int{d^3 x}      
\left\{B{\pa}^mA_m+ {\x\over2}B^2 + {\ov c}\Box c \right\} 
\;\;\;\;\;\;\;,\label{gf}
\eq
\bq
\S_{\rm ext}\!\!\!\!&=&\!\!\!\!\int{d^3 x}      
\left\{ \ov\Ome_+s\j_+ - \ov\Ome_-s\j_- -
s\ov\j_+\Ome_+ + s\ov\j_-\Ome_- + s\,\q\,\Q + sH\,\X \right\} \;\;\;\;\;\;\;.
\label{ext} 
\eq

The BRS transformations are given by :
\bq
&&sH=-Qc\q\;\;,\;\;\; s\q=Qc(v+H)\;\;, \nonumber\\
&&s\j_{\pm}=iqc\j_{\pm} \;\;,\;\;\; s\ov\j_{\pm}=-iqc\ov\j_{\pm}\;\;,
\nonumber\\
&&sA_m=-{1\over g}{\pa}_m c \;\;,\;\;\; sc=0 \;\;,\nonumber\\
&&s{\ov c}={1\over g}B \;\;,\;\;\; sB=0 \;\;,
\eq
where $c$ is the ghost, ${\ov c}$ is the anti-ghost and $B$ is the 
Lagrange multiplier field.

The complete action, $\S$, we are considering here is
\bq
\S=\S_{\rm inv}+\S_{\rm gf}+\S_{\rm ext}\;\;\;.\label{total}
\eq

The QED$_{3}$-action
(\ref{total}) is invariant under the reflexion 
symmetry, $P$, whose action on the fields and external sources is 
fixed as below :
\begin{equation}\begin{array}{lll}
x_m &\ \pari\ & x_m^P=(x_0,-x_1,x_2)\;\;\;, \\
\j_{\pm} &\ \pari\ & \j_{\pm}^P=-i\g^1\j_{\mp}\;\;,\qquad 
\ov\j_{\pm}~\ \pari~\ \ov\j_{\pm}^P=i\ov\j_{\mp}\g^1\;\;\;, \\
A_m &\ \pari\ & A_m^P=(A_0,-A_1,A_2)\;\;\;,\\  
{\f} &\ \pari\ & {\f}^P=\f\ ,\qquad \f= H,\,\q,\,c,\,\bar c,\,B \;\;\;, \\
\Ome_{\pm} &\ \pari\ & \Ome_{\pm}^P=-i\g^1\Ome_{\mp} \;\;,\qquad 
\ov\Ome_{\pm}~\ \pari\ ~\ov\Ome_{\pm}^P=i\ov\Ome_{\mp}\g^1\;\;\;,\\
\Q &\ \pari\ & \Q^P=\Q \;\;,\qquad \X ~\ \pari~\  \X^P=\X
\;\;\;.
\end{array}\label{xp}\end{equation}
The ultraviolet and infrared dimensions, $d$ and $r$ respectively, 
as well as the ghost numbers, $\F\Pi$, and the Grassmann parity, $GP$, 
of all fields and sources are collected in Table~\ref{table1}.

\begin{table}[hbt]
\centering
\begin{tabular}{|c||c|c|c|c|c|c|c|c|c|c|}
\hline
    &$A_m$ &$H$ &$\q$ &$\j_{\pm}$ &$c$ &${\ov c}$ &$B$ &$\Q$ &$\X$ 
    &$\Ome_{\pm}$  \\
\hline\hline
$d$ &${1\over2}$ &${1\over2}$ &${1\over2}$ &1 &0 &1 &${3\over2}$ 
&${5\over2}$ &${5\over2}$ &2  \\
\hline
$r$ &${3\over2}$ &${3\over2}$ &${1\over2}$ &${3\over2}$ &0 &3 
&${3\over2}$ &${5\over2}$ &${5\over2}$ &$2$  \\
\hline
$\F\Pi$&0 &0 &0 &0 &1 &$-1$ &0 &$-1$ &$-1$ &$-1$  \\
\hline
$GP$&0 &0 &0 &1 &1 &1 &0 &1 &1 &0  \\
\hline
\end{tabular}
\caption[t1]{UV and IR dimensions, $d$ and $r$, ghost numbers, 
$\F\Pi$, and Grassmann parity, $GP$.}
\label{table1}
\end{table}

The BRS invariance of the action is expressed in a 
functional way by the Slavnov-Taylor identity
\bq
\cs(\S)\!\!\!\!&=&\!\!\!\!\int{d^3 x}      
\left\{-{1\over g}{\pa}^m c {\d\S\over\d A^m} + 
{1\over g}B {\d\S\over\d {\ov c}} +
{\d\S\over\d \ov\Ome_+}{\d\S\over\d \j_+} -
{\d\S\over\d \ov\Ome_-}{\d\S\over\d \j_-} -
{\d\S\over\d \Ome_+}{\d\S\over\d \ov\j_+} +
{\d\S\over\d \Ome_-}{\d\S\over\d \ov\j_-} \;+\right.   
\nonumber\\
&&\left.
-\;{\d\S\over\d\Q }{\d\S\over\d\q } - 
{\d\S\over\d\X}{\d\S\over\d H}
\right\}=0 \;\;\;\;\;\;\;.\label{slavnov} 
\eq

In addition to the Slavnov-Taylor identity (\ref{slavnov}) 
the gauge condition, the ghost equation and the antighost equation 
read
\letra
\bq
{\d\S\over\d B}\!\!\!\!&=&\!\!\!\!{\pa}^mA_m + \x B \;\;\;;
\label{gaugecond}\\
{\d\S \over\d \ov c}\!\!\!\!&=&\!\!\!\!\Box\,c  \;\;\;;\label{ghostcond}\\
-i{\d\S \over\d c}\!\!\!\!&=&\!\!\!\!i \Box\,{\ov c} -i
{\d\S_{\rm ext}\over \d c}  
\;\;\;.\label{antighostcond}
\eq
\antiletra

We notice that, the right-hand sides of 
Eqs.(\ref{gaugecond} -- \ref{antighostcond}) are linear in the 
quantum fields, then are not subjected to renormalizations.

The solution for the Eqs.(\ref{gaugecond} -- \ref{ghostcond}) 
is simply
\be
\S=\bar\S(\j_{\pm},H,\q,A_m,c,\Ome_{\pm},\Q,\X)+\int{d^3 x}      
\left\{B{\pa}^mA_m + {\x\over2}B^2 + {\ov c}\Box c \right\}\;\;\;\;. 
\label{landauaction}
\ee
Putting these informations into (\ref{slavnov}) we find that the 
constraint on $\bar\S$ is given by
\bq
{\bar\cs}(\bar\S)\!\!\!\!&=&\!\!\!\!\int{d^3 x}      
\left\{-{1\over g}{\pa}^m c {\d\bar\S\over\d A^m} + 
{\d\bar\S\over\d \ov\Ome_+}{\d\bar\S\over\d \j_+} -
{\d\bar\S\over\d \ov\Ome_-}{\d\bar\S\over\d \j_-} -
{\d\bar\S\over\d \Ome_+}{\d\bar\S\over\d \ov\j_+} +
{\d\bar\S\over\d \Ome_-}{\d\bar\S\over\d \ov\j_-} \;+\right.   
\nonumber\\
&&\left.
-\;{\d\bar\S\over\d\Q }{\d\bar\S\over\d\q } - 
{\d\bar\S\over\d\X}{\d\bar\S\over\d H}
\right\}=0 \;\;\;\;\;\;\;.\label{barslavnov} 
\eq
The corresponding linearized Slavnov-Taylor operator for any functional 
$\bar\cf$ reads
\bq
{\bar\cs}_{\bar\cf}\!\!\!\!&=&\!\!\!\!\int{d^3 x}      
\left\{-{1\over g}{\pa}^m c {\d\over\d A^m} + 
{\d\bar\cf\over\d \ov\Ome_+}{\d\over\d \j_+} -
{\d\bar\cf\over\d \ov\Ome_-}{\d\over\d \j_-} +
{\d\bar\cf\over\d \j_+}{\d\over\d \ov\Ome_+} -
{\d\bar\cf\over\d \j_-}{\d\over\d \ov\Ome_-} \;+\right.   \nonumber\\
&&\left. 
-\;{\d\bar\cf\over\d \Ome_+}{\d\over\d \ov\j_+} +
{\d\bar\cf\over\d \Ome_-}{\d\over\d \ov\j_-} -
{\d\bar\cf\over\d \ov\j_+}{\d\over\d \Ome_+} +
{\d\bar\cf\over\d \ov\j_-}{\d\over\d \Ome_-} +
{\d\bar\cf\over\d\Q }{\d\over\d\q } +
{\d\bar\cf\over\d\q }{\d\over\d\Q } \;+\right.   
\nonumber\\
&&\left.
-\;{\d\bar\cf\over\d\X }{\d\over\d H } -
{\d\bar\cf\over\d H }{\d\over\d\X } 
\right\} \;\;\;\;\;\;\;.\label{slavnovlin} 
\eq
The following nilpotency identities holds :
\letra
\bq
{\bar\cs}_{\bar\cf}{\bar\cs}(\bar\cf)=0 \;\;,\;\;\;\forall\;
   \bar\cf \;\;\;;\label{nilpot1} \\
{\bar\cs}_{\bar\cf}{\bar\cs}_{\bar\cf}=0 \;\;\;{\mbox{if}} \;\;\;
   {\bar\cs}(\bar\cf)=0\;\;\;. \label{nilpot2}
\eq
\antiletra
The operation of $\bar\cs_{\bar\S}$ 
over the fields and the external sources is 
given by
\bq
&&\bar\cs_{\bar\S}\f=s\f \;\;,\;\;\;\f=\j_{\pm},\;\ov\j_{\pm},\;
H,\;\q,\;A_m,\;c,\;
{\ov c}\!\!\aand\!\! B \;\;\;,\nonumber\\
&&\bar\cs_{\bar\S}\ov\Ome_+={\d\S\over\d \j_+} \;\;,\;\;\;
\bar\cs_{\bar\S}\ov\Ome_-=-{\d\S\over\d \j_-} 
\;\;\;,\nonumber\\
&&\bar\cs_{\bar\S}\Ome_+=-{\d\S\over\d \ov\j_+}  \;\;,\;\;\;  
\bar\cs_{\bar\S}\Ome_-={\d\S\over\d \ov\j_-} \;\;\;,\nonumber\\
&&\bar\cs_{\bar\S}\Q=-{\d\S\over\d\q }  \;\;,\;\;\; 
\bar\cs_{\bar\S}\X=-{\d\S\over\d H }\;\;\;.\label{operation1}
\eq
In order to study the stability~{\cite{brs}} of the action (\ref{total}) 
under the radiative corrections, we perturb the classical action by local 
functional, $\S^c$, having the same quantum numbers as $\bar\S$ :
\bq
\bar\S \longrightarrow \bar\S'=\bar\S + \e\S^c 
\eq
where $\e$ is an infinitesimal parameter.
Then requiring that the perturbed action $\S'$ satisfies the same 
conditions as $\bar\S$ we obtain :
\bq
{\bar\cs}_{\bar\S}\S^c=0 \;\;\;\;, \label{stabcond}
\eq  
\bq
{\d\S^c\over{\d B} }=0 \;\;,\;\;\;{\d\S^c\over{\d{\ol c}}}=0 
\;\;,\;\;\; {\d\S^c\over{\d c}}=0 \;\;\;\;, \label{suplcond}
\eq
which follow from the Slavnov-Taylor identity and from the conditions 
(\ref{gaugecond} -- \ref{antighostcond}), and, moreover:
\bq
W_{\rm rigid} {\S}^{\,c}=0\;\;\;\;, \label{crigidcond}
\eq
where $W_{\rm rigid}$ is the Ward operator of rigid $U(1)$ symmetry 
defined 
by
\bq
W_{\rm rigid}\!\!\!\!&=&\!\!\!\!\int{d^3 x}      
\left\{
q\j_+{\d\over\d \j_+} +
q\j_-{\d\over\d \j_-} -
q\ov\j_+{\d\over\d \ov\j_+} -
q\ov\j_-{\d\over\d \ov\j_-} -
iQ(v+H){\d\over\d\q} +
iQ\q{\d\over\d H} \;+\right.   
\nonumber\\
&&\left.
+\;q\Ome_+{\d\over\d \Ome_+} +
q\Ome_-{\d\over\d \Ome_-} -
q\ov\Ome_+{\d\over\d \ov\Ome_+} -
q\ov\Ome_-{\d\over\d \ov\Ome_-} -
iQ\X{\d\over\d\Q} +
iQ\Q{\d\over\d\X}
\right\}\;\;\;\;\;. \label{wrigid} 
\eq
Eq.(\ref{crigidcond}) follows 
from the rigid $U(1)$ invariance of the 
action\footnote{ Rigid invariance itself 
follows from the antighost equation
(\ref{antighostcond}) and from the validity of the Slavnov-Taylor 
identity (\ref{slavnov}).} in the Landau gauge:
\bq
W_{\rm rigid} \S=0 \;\;\;\;. \label{rigidcond}
\eq

The BRS consistency condition in the ghost number sector zero, given 
by Eq.(\ref{stabcond}), constitutes a cohomology problem due to the 
nilpotency (\ref{nilpot2}) of the linearized Slavnov-Taylor operator 
(\ref{slavnovlin}). Its solution can always be written as a sum 
of a trivial cocycle $\bar\cs_{\bar\S}\wh\S$, where $\wh\S$ has 
ghost number $-1$, and nontrivial elements $\S_{\rm phys}$
belonging to the cohomology 
of $\bar\cs_{\bar\S}$ (\ref{slavnovlin}) in the sector of
ghost number zero: 
\be
\S^c=\S_{\rm phys} + \bar\cs_{\bar\S}\wh\S \;\;\;, \label{split}
\ee
where the trivial cocycle $\bar\cs_{\bar\S}\wh\S$ corresponds to field
renormalizations, which are unphysical. On the other hand, the 
nontrivial perturbation $\S_{\rm phys}$ 
corresponds to a redefinition of the physical parameters -- 
coupling constants and masses. An explicit computation, yields
the following solution for Eq.(\ref{split}):
\letra
\bq
\S_{\rm phys} \!\!\!\!&=&\!\!\!\! z_g\left(g{\pa\over{\pa g}} - N_A + N_B -
2\x{\pa\over{\pa \x}}\right)\S + z_m\;m{\pa\S\over{\pa m}}\;+ \nonumber\\
&&+\; z_y\;y{\pa\S\over{\pa y}} + z_{M^2_H}\;M^2_H{\pa\S\over{\pa M^2_H}} + 
 z_{\z}\;\z{\pa\S\over{\pa \z}} + z_{\l}\;\l{\pa\S\over{\pa \l}}  \;\;\;,\\
\bar\cs_{\bar\S}\wh\S \!\!\!\!&=&\!\!\!\!  \bar\cs_{\bar\S} \int{d^3 x} 
\left\{ 
  z_{\j}\left( \ov\j_+\Ome_+ - \ov\Ome_+\j_+ -  \ov\j_-\Ome_- 
  + \ov\Ome_-\j_- \right) + z_H\left[\q\Q+(v+H)\X \right]
     \right\}  \nonumber\\
\!\!\!\!&=&\!\!\!\! z_{\j}\left( N_{\j_+} + N_{\ov\j_+} +N_{\j_-} 
  + N_{\ov\j_-} - N_{\Ome_+} - N_{\ov\Ome_+} - N_{\Ome_-} 
  - N_{\ov\Ome_-} \right)\S \nonumber\\ 
&&+\;z_H\left( N_\q+{\wh N}_H-N_\Q-N_\X \right) \S
  \;\;\;, \label{countcount}
\eq
\antiletra
where the counting operators are defined by 
\bq
N_{\f}\!\!\!\!&=&\!\!\!\!\int{d^3 x} \;\f\;{\d\over\d \f} \;\;,\;\;\;\f=
 \j_{\pm},\;
\ov\j_{\pm},\;\q,\;\Ome_{\pm},\;\ov\Ome_{\pm},\;\Theta,\; \Xi,\;
\;A_m\!\!\aand\!\! B \;\;\;, \nonumber\\
 \wh N_H \!\!\!\!&=&\!\!\!\! \int{d^3 x}\;(v+H)\;{\d\over\d H}
\;\;\;.
\eq
The stability proof we have given corresponds, at the quantum level, to
the multiplicative renormalizability of the model: 
all the possible counterterms induced by
the radiative corrections correspond to a redefinition of the parameters
of the starting classical theory. The parameters 
$z_g$, $z_{m}$, $z_y$, $z_{M_H^2}$, $z_{\z}$, 
$z_{\l}$, $z_H$ and $z_{\j}$  are then renormalization constants,
which are fixed by the following normalization conditions -- expressed
on the vertex functional $\G$, which coincides with the classical action
$\S$ in the classical limit: 
\bq
&&\G_{HH}(p^2)\bigg|_{p^2=M_H^2}=0 \;\;,\;\;\;{\pa\over\pa p^2}
\G_{HH}(p^2)\bigg|_{p^2=\k^2}=1 \;\;,\nonumber\\
&&\G_{HHHH}(p)\bigg|_{p=\bar p(\k)}=-\z\;\;,\;\;\;
 \G_{HHHHHH}(p)\bigg|_{p=\bar p(\k)}=-\l\;\;,\nonumber\\
&&\G_{\ol\j_{\pm}\j_{\pm}}(\sl p)\bigg|_{{\ssl p}=\pm m}=0 \;\;,
\;\;\;{\pa\over{\pa\sl p}}\G_{\ol\j_{\pm}\j_{\pm}}(\sl p)
\bigg|_{{\ssl p}=\k}=1 \;\;,\nonumber\\
&&{\pa\over\pa p^2}\G_{A^TA^T}(p^2)\bigg|_{p^2=\k^2}=1
 \;\;,\;\;\;\G_{{\ol\j_{\pm}}\j_{\pm} HH}(p)
\bigg|_{p=\bar p(\k)}=\mp y \;\;\;, \label{normcond}
\eq
where $\k$ is an energy scale and $\bar p(\k)$ some reference set of 
4-momenta at this scale. 

To complete the proof of the renormalizability of the model, we show 
that all the symmetries defining the model can be extended to the 
quantum level, for the vertex functional $\G$
\be
\G=\S + {\co}(\hbar) \;\;\;\;.\label{vertex}
\ee
Now, it is trivial to verify that the solution of 
Eqs.(\ref{gaugecond} -- \ref{ghostcond}), that are linear in 
the quantized fields, is given by
\be
\G=\bar\G(\j_{\pm},H,\q,A_m,c,\Ome_{\pm},\Q,\X)+\int{d^3 x}      
\left\{B{\pa}^mA_m + {\x\over2}B^2 + {\ov c}\Box c \right\} 
\label{glandauaction}
\ee
As a consequence we have the following conditions on $\bar\G$ -- 
defined from $\G$ similarly to (\ref{landauaction}): 
\letra
\bq
&&{\d\bar\G\over\d B}=0\;\;\;;\label{qgaugecond}\\
&&{\d\bar\G\over\d\ol c}=0 \;\;\;;\label{qghost}\\
&&-i{\d\bar\G\over\d c}=-i{\d\S_{\rm ext}\over\d c}\;\;\;;
\label{qantighost}\\
&&W_{\rm rigid} \bar\G=0 \;\;\;, \label{qrsuplcond}
\eq
\antiletra
where $W_{\rm rigid}$ has already been defined by (\ref{wrigid}),
and where (\ref{qantighost}) is the quantum extension of 
Eq.(\ref{antighostcond}).  

According to the Quantum Action Principle~{\cite{sopi,qap}} the 
Slavnov-Taylor identity (\ref{slavnov}) gets a quantum breaking
\be
\cs(\G)=\bar\cs(\bar\G)=\D \cdot \G = \D + {\co}(\hbar \D)\;\;\;, 
\label{slavnovbreak}
\ee
where $\D$ is a local integrated functional with ghost number one.

The nilpotency identity ({\ref{nilpot1}) together with
\be
{\bar\cs}_{\bar\G}={\bar\cs}_{\bar\S} + {\co}(\hbar)
\ee
implies the following consistency condition for the breaking $\D$ :
\be
{\bar\cs}_{\bar\S}\D=0 \;\;\;.\label{breakcond1}
\ee

In order to identify other constraints for $\D$, 
we use the following algebraic relations, valid for any functional 
$\bar\cf$ with zero $GP$: 
\letra
\bq
&&{\d\bar\cs({\bar\cf})\over\d B} - 
\bar\cs_{{\bar\cf}}{\d{\bar\cf}\over\d B}= 0 \;\;\;;\label{fcond1}\\
&&{\d\bar\cs({\bar\cf})\over\d\ol c} + 
\bar\cs_{{\bar\cf}}{\d{\bar\cf}\over\d\ol c }=0\;\;\;;\label{fcond2} \\
&&-i\int{d^3 x}{\d\over\d c}\bar\cs({\bar\cf}) + 
\bar\cs_{{\bar\cf}}\int{d^3 x}\left(-i{\d\over\d c}{\bar\cf}
 +i{\d\S_{\rm ext}\over\d c}\right)
=W_{\rm rigid} \bar\cf \;\;\;; \label{fcond3}\\
&&W_{\rm rigid}\bar\cs(\bar\cf) - \bar\cs_{\bar\cf}W_{\rm rigid} \bar\cf =0 
\;\;\;.\label{fcond4}
\eq
\antiletra
Taking into account Eqs.(\ref{qgaugecond} -- \ref{qrsuplcond}), 
Eq.(\ref{slavnovbreak}) and assuming $\bar\cf$$=$$\bar\G$ in 
Eqs.(\ref{fcond1} -- \ref{fcond4}), the following consistency 
conditions on the breaking $\D$ are found :
\letra
\bq
&&{\d\D\over\d B}=0  \;\;\;;\label{breakcond2}\\
&&{\d\D\over\d\ol c}=0 \;\;\;;\label{breakcond3}\\
&&\int{d^3 x}{\d\over\d c}\D=0 \;\;\;;\label{breakcond4}\\[3mm]
&&W_{\rm rigid} \D=0 \;\;\;.\label{breakcond5}
\eq
\antiletra
The Wess-Zumino consistency condition (\ref{breakcond1}) constitutes a 
cohomology problem like in the zero ghost number case (\ref{stabcond}).
Its solution can always be written as a sum of a trivial cocycle 
$\cs_{\S}{\wh\D}^{(0)}$, where ${\wh\D}^{(0)}$ 
has ghost number $0$, and of nontrivial elements belonging to the 
cohomology of $\cs_{\S}$ (\ref{slavnovlin}) in the sector of ghost number 
one:
\be
\D^{(1)} = {\wh\D}^{(1)} + \cs_{\S}{\wh\D}^{(0)} \;\;\;, 
\label{breaksplit}
\ee
where $\D^{(1)}$ must be even under $P$-symmetry and obey the conditions 
imposed by Eqs.(\ref{breakcond2} -- \ref{breakcond5}). 
The trivial cocycle $\cs_{\S}{\wh\D}^{(0)}$ can be absorbed into the
 vertex functional $\G$ as a local integrated noninvariant 
couterterm $-{\wh\D}^{(0)}$. 

Now, from the condition (\ref{breakcond4}), we conclude that  
\be
\D^{(1)} = \int{d^3 x} \;K^{(0)}_m\;{\pa}^mc \;\;\;. \label{anomaly}
\ee
By analyzing the Slavnov-Taylor operator $\bar\cs_{\bar\S}$ 
(\ref{slavnovlin}) and the Wess-Zumino consistency condition 
(\ref{breakcond1}), we see that the UV and IR dimensions of
the breaking $\D^{(1)}$ are bounded by 
$d$$\leq$${7\over2}$ and $r$$\geq$$2$. Therefore, $K^{(0)}_m$, of 
ghost number $0$,
has dimensions bounded by 
$d$$\leq$${5\over2}$, $r$$\geq$$1$.
 
Now, rewriting $K^{(0)}_m$ as a linear combination
\be
K^{(0)}_m = {\sum_{i=1}^{8} }\; a_i \; K^{(0)i}_m \;\;\;, \label{lincombk0}
\ee
where
\bq
&&K^{(0)1}_m = A_m \;\;,\;\;K^{(0)2}_m = A_m A^nA_n\;\;, \nonumber\\
&&K^{(0)3}_m = A_m (A^nA_n)^2 \;\;,\;\;
K^{(0)4}_m = A_m (\ov\j_+\j_+ - \ov\j_-\j_-)\;\;, \nonumber\\ 
&&K^{(0)5}_m = A_m A^nA_n ((v+H)^2+\q^2) \;\;,\;\; 
K^{(0)6}_m = A_m ((v+H)^2+\q^2) \;\;,\nonumber\\
&&K^{(0)7}_m = A_m ((v+H)^2+\q^2)^2 \aand
K^{(0)8}_m = \ov\j_+\g_m\j_+ + \ov\j_-\g_m\j_- \;\;\;,  
\eq
and solving all the conditions it has to fulfil, we 
can easily show, with the help of Eqs.(\ref{operation1}), that there 
exist local functionals ${\wh\D}^{(0)i}$ such that
\be 
\int d^3x \;K^{(0)i}_m\;\pa^m c = \cs_{\S}{\wh\D}^{(0)i}\ ,\ \ 
     i=1,\cdots,8\;\;\:.
\label{lincomd0}
\ee
This means that ${\wh\D}^{(1)}=0$ in (\ref{breaksplit}), which
implies the implementability of the Slavnov-Taylor identity to every
order through the absorbtion of the noninvariant counterterm 
$-\sum_ia_i{\wh\D}^{(0)i}$.

In conclusion, the algebraic method of renormalization allowed us 
to show that the model is perturbatively renormalizable to all orders. 
The study of the possible counterterms has led to the conclusion 
that the 
model is multiplicatively renormalizable, namely that the 
counterterms can be reabsorbed by a redefinition of the initial 
parameters. Finally, we have proven that anomalies are absent. 
We stress that our algebraic analysis does not involve any 
regularization scheme, nor any particular diagramatic calculation.

\small

\subsection*{Acknowledgements}
The authors express their gratitude to Prof. S.P. Sorella for 
patient and helpful discussions. One of 
the authors (O.M.D.C.) dedicates 
this work to his wife,
Zilda Cristina, and to his daughter, Vittoria, who was 
born in 20 March 1996. O.M.D.C. and D.H.T.F. thank to the 
{\it High Energy Section} of the {\it ICTP - Trieste - Italy}, 
where this work was done, for the kind hospitality and 
financial support, and to its Head, Prof. S. Randjbar-Daemi. O.P. 
thanks the CBPF, the Physics Department of the Federal University 
of Esp\'\i rito Santo (Brazil) and his collegues from both institutions 
for their very kind hospitality.
Thanks are also due to the Head of CFC-CBPF,
Prof. A.O. Caride, and Dr. R. Paunov, for encouragement. 
CNPq-Brazil is 
acknowledged for
invaluable
financial help.



\begin{thebibliography}{99}
\bibitem{e-pair}{M.A. De Andrade, O.M. Del Cima and J.A. Helay\"el-Neto, 
{\em{Electron-pair condensation in parity-preserving QED$_3$}}, 
hep-th/9603054. Talk given at {\it Quantum Systems: New Trends 
and Methods 96 - QS96}, Minsk, Belarus.}
\bibitem{brs}{C. Becchi, A. Rouet and R. Stora, {\em{Comm. Math. Phys.}} 
{\bf{42}} (1975) 127 and {\em{Ann. Phys. (N.Y.)}} {\bf{98}} (1976) 287 ; 
O. Piguet and A. Rouet, {\em{Phys. Rep.}} {\bf{76}} (1981) 1.}
\bibitem{troisieme}{O. Piguet, {\em{Renormalisation en th\'eorie quantique 
des champs}} and {\em{Renormalisation des th\'eories de jauge}}, 
lectures of the {\em{Troisi\`eme cycle de la physique en Suisse Romande}} 
(1982-1983).}
\bibitem{sopi}{O. Piguet and S.P. Sorella, {\em{Algebraic Renormalization}}, 
Lecture Notes in Physics, m28, Springer-Verlag, Berlin, Heidelberg, 1995.}
\bibitem{qap}{J.H. Lowenstein, {\em{Phys. Rev.}} {\bf{D4}}
 (1971) 2281 and {\em{Comm. Math. Phys.}} {\bf{24}} (1971) 1 ; Y.M.P. Lam, 
{\em{Phys. Rev.}} {\bf{D6}}
 (1972) 2145 and {\em{Phys. Rev.}} {\bf{D7}} (1973) 2943 ; T.E. Clark and 
J.H. Lowenstein, {\em{Nucl. Phys.}} {\bf{B113}} (1976) 109.}
\bibitem{arqed31}{O.M. Del Cima, D.H.T. Franco, J.A. Helay\"el-Neto and 
O.Piguet, {\em Algebraic renormalization of parity-preserving QED$_{3}$ 
coupled to scalar matter I: unbroken case}, 
UGVA-DPT-1996-09-950.}
\bibitem{binegar}{B. Binegar, {\em{J. Math. Phys.}} 
{\bf{23}} (1982) 1511.}
\bibitem {abers}{E.S. Abers and B.W. Lee,  
 {\em Phys. Rep.} {\bf 9} (1973) 1.}
\bibitem{gaugeind}{O. Piguet and K. Sibold, {\em{Nucl. Phys.}} 
{\bf{B253}} (1985) 517.} 


\end{thebibliography}
\end{document}